\documentclass[twocolumn,showpacs,preprintnumbers,amsmath,amssymb]{revtex4}
\usepackage{graphicx,psfrag}
\begin{document}
\title{ Absorption cross section of RN black hole }
\author{Sini R}
\email {sini@cusat.ac.in}
\author{V C Kuriakose}
\email {vck@cusat.ac.in} \affiliation{Department of Physics, Cochin
University of Science and Technology, Kochi 682022, India.}
\begin{abstract}
The behavior of a charged scalar field in the RN black hole space
time is studied using WKB approximation. In the present work it is
assumed that matter waves can get reflected from the event horizon.
Using this effect, the Hawking temperature and the absorption cross
section for RN black hole placed in a charged scalar field are
calculated.  The absorption cross section $\sigma _{abs}$ is found
to be inversely proportional to square of the Hawking temperature of
the black hole.
\end{abstract}
 \maketitle

\section{Introduction}
     Black holes are natural, stable solutions of Einstein equations in general relativity.
A black hole is distinguished by the fact that no information can
escape from within the event horizon. The presence of black holes
can be inferred only through indirect methods. One of the most
useful and efficient ways to study the properties of black hole is
by scattering matter waves off them \cite{ss}. By studying how a
black hole interact with its environment one may understand whether
there are ways of making more direct observations of such, in
principle, invisible objects. These studies will be useful for
understanding of the signals expected to be received by the new
generation of gravitational-wave detectors in the near future
\cite{tt} which is certainly one of the most challenging tasks for
modern observational astronomy.

    A considerable effort has taken place in studying the waves
scattered off by black holes. Both numerical and analytical methods
in solving the various wave equations in black hole scattering have
been developed \cite{ss,ii}. In classical general relativity black
hole's properties can be precisely calculated and the black holes
may be thought of as astronomical objects with masses about several
times of sun. Hawking found when the laws of quantum field theory is
applied black holes are not truly black. It possesses entropy and
temperature and quantum mechanically a black hole with temperature
is able to emit radiation, leading to a situation of particle
production int the presence of black holes. When we try to quantize
gravitational force, we have to consider how does quantum mechanics
affect the behavior of black holes.

Interest in the absorption of quantum waves by black hole was
reignited in the 1970s, following Hawking's discovery that black
holes can emit, as well as scatter and absorb, radiation
\cite{abaa}. Hawking showed that the evaporation rate is
proportional to the total absorption cross section. Unruh \cite{pp}
found the absorption cross section for massive scalar and Dirac
particles scattered off by small non rotating black holes. In a
series of papers \cite{abab,abac,abad} Sanchez considered the
scattering and absorption of massless scalar particles by an
uncharged, spherically symmetric black hole.

 Another quantum effect of intrest is that event horizons need not be fully absorptive
type but can reflect waves falling on it. It is also proposed that
event horizons has a finite energy width. 't Hooft \cite{viv}
explained the horizon of the black hole as a brick wall so that the
outer horizon $r_+$ spreads into a range of
$\left(r_+-\Delta,r_++\Delta\right)$. Quantum horizon concepts were
introduced by Mu-Lin Yan and Hua Bai \cite{22}. The relevant
equation governing a scattering process in a black hole space time
is analogous to Schrodinger type equations governing scattering
phenomena in quantum mechanics. Hence the standard techniques used
to study quantum scattering can be used to study scattering problems
in black hole space time.

 In the present work we study the scattering of charged scalar waves in the
Reissner Nordstorm (RN) space-time. Earlier several authors have
studied scattering of scalar and Fermi fields under different black
hole space time and calculated absorption cross sections. In all
these calculations the black hole is assumed to be capable of
absorbing the radiation falling on it, but here we consider that
both absorption and reflection could take place at the horizon of
black holes. Kurchiev \cite{bb} has also calculated the absorption
coefficient of scalar waves in Schwarzschild space time using the
phenomenon of reflection of waves at the event horizon. The
absorption coefficient of scalar waves in Schwarzschild de Sitter
space time was found earlier \cite{sr}. In this work we use WKB
approximation, which has been proven to be useful in many cases such
as,  the evaluation of the quasinormal mode frequencies
\cite{uu,vv}, finding the solution of wave equation in the vicinity
of event horizon of black holes, etc. In section II we explain the
nature of radial wave functions in different regions of RN space
time. Section III contains calculation of absorption cross section
for charged scalar wave scattered off by RN black hole, wherein we
take into consider both reflection and absorption properties of the
black hole horizon. Section IV concludes the paper.
\section{Nature of radial Wave functions in different regions of RN space time}
   The metric describing a charged spherical symmetric black hole, written
in spherical polar coordinates, is given by
\begin{eqnarray}
\label{a11}ds^2=\left( 1-\frac 1r+\frac{q^2}{r^2}\right) dt^2-\frac
1{\left( 1-\frac 1r+\frac{q^2}{r^2}\right) }dr^2 \\
\nonumber-r^2\left( d\theta ^2+\sin ^2\theta d\phi ^2\right).
\end{eqnarray}
The outer and inner horizons of the RN black hole are,
\begin{equation}
r_{\pm }=\frac 12\pm \sqrt{\frac 14-q^2}.
\end{equation}
The dynamical behavior of a massive charged scalar field under RN
background is \cite{bn},
\begin{equation}
\Psi _{;ab}g^{ab}+\imath eA_ag^{ab}[2\Psi _{;b}+\imath eA_b\Psi
]+\imath eA_{a;b}g^{ab}\Psi +\mu ^2\Psi =0.
\end{equation}
The radial part is separated out by putting, $\Psi =\exp
(-\imath\epsilon t)\Phi _l\left( r\right) Y_{lm}\left( \theta ,\phi
\right)$,
where $\epsilon $, $l$, and $m$ are energy, momentum and its projection, while $%
\Phi _l(r)$ is radial function, then we will obtain

\begin{eqnarray}
\label{g6}\Delta \Phi _l^{\prime \prime }(r)+(2r-1)\Phi _l^{\prime
}(r)&+&\\\nonumber\left( \frac{\left( \epsilon -eA_t\right)
^2r^4}{\Delta ^{}}-\mu ^2r^2-l\left( l+1\right) \right) \Phi
_l(r)=0,
\end{eqnarray}
where $e$ is the electric charge, $A_t=\frac qr$ is the electric
potential and $\Delta =r^2-r+q^2=\left( r-r_{+}\right) \left(
r-r_{-}\right)$. Therefore Eq. (\ref{g6}) can be written as
\begin{eqnarray}
\label{h6}\left( r^2-r+q^2\right) \Phi _l^{\prime \prime
}(r)+(2r-1)\Phi
_l^{\prime }(r)&+&\\ \nonumber\left( \frac{\left( \epsilon -eA_t\right) ^2r^4}{r^2-r+q^2}%
-\mu ^2r^2-l\left( l+1\right) \right) \Phi _l(r)=0.
\end{eqnarray}
The radial equation can also be written as
\begin{eqnarray}
\frac d{dr}(r^2-r+q^2)\Phi _l^{\prime }(r)&+&\\ \nonumber\left(
\frac{\left( \epsilon -eA_t\right) ^2r^2}{\left( 1-\frac
1r+\frac{q^2}{r^2}\right) }-\mu ^2r^2-l\left( l+1\right) \right)
\Phi _l(r)=0.
\end{eqnarray}
To study the scattering problem we divide the space time into 3
regions \cite{pp}. We consider the three different regions starting
from the event horizon as shown below.

\subsection{Region 1: Vicinity of horizon: $r\rightarrow r_{+}$}

We solve the wave equation near the horizon and also evaluate
Hawking temperature using WKB approximation. By using WKB
approximation $\Phi =\exp ^{-\imath \int k(r)dr}$ in Eq. (\ref{h6}),
and equating the real part we will get the radial wave number
$k(r,l,\varepsilon )$ from the corresponding equation of motion:
\begin{eqnarray}
\label{14}k^2(r)=\left( 1-\frac 1r+\frac{q^2}{r^2}\right) ^{-1}\\ \nonumber\left[ \frac{%
\left( \epsilon -eA_t\right) ^2}{\left( 1-\frac 1r+\frac{q^2}{r^2}\right) }-%
\frac{l(l+1)}{r^2}-\mu ^2\right] ,
\end{eqnarray}
which gives,
\begin{eqnarray}
\label{c41}k\left( r\right) =\pm \left[ \left( \epsilon -eA_t\right)
^2r^4-\left( l(l+1)+\mu ^2r^2\right) (r^2-r+q^2)\right] ^{\frac
12}\\ \nonumber\frac 1{(r^2-r+q^2)}.
\end{eqnarray}

Thus near the outer horizon $r\rightarrow r_{+}$, we will get $
k\left( r\rightarrow r_{+}\right) =\pm \frac \xi {\left(
r-r_{+}\right) } $, where $ \xi =\frac{\left( \epsilon -eA_t\right)
r_{+}^2}{\left( r_{+}-r_{-}\right) }$. Therefore the wave function
in the region $r\rightarrow r_{+}$ can be written as,
\begin{equation}
\label{lm}\Phi _l(r)\sim\exp (\pm \imath \int \frac \xi {\left(
r-r_{+}\right) }dr)=\exp \left( \pm \imath \xi \ln \left(
r-r_{+}\right) \right),
\end{equation}
i.e.,
\begin{equation}
\label{p}\Phi _l(r) \sim\exp \left( \pm \imath \xi \ln \left(
r-r_{+}\right) \right).
\end{equation}
Let us describe its radial motion with the help of the wave function
$\Phi \left( r\right) $. Using Eq. (\ref{lm}) the wave function in
the vicinity of horizon can be written, assuming that the wave gets
reflected at the horizon, as
\begin{equation}
\label{p1}\Phi_{l} \left( r\right) \sim\exp \left( - \imath \xi \ln
\left( r-r_{+}\right) \right) +\mid R\mid \exp \left( + \imath \xi
\ln \left( r-r_{+}\right) \right),
\end{equation}
where $R$ represents the reflection coefficient and the solution
represents the interference between the incident  and reflected
waves. If $R\neq 0$, there is a definite probability for the
incident waves to get reflected at the horizon. The wave function is
singular at $r=r_{+}$. We now consider a point distant
$z=r-r_{+}\,$, from the horizon and treat $z$ as a complex variable.
The above wave function is analytic in $z,$ except for the power
type singularity at $z=0$ which induces a cut emerging from this
point on the complex plane $z$. Let us take $r$ outside the outer
horizon region but close to the vicinity of black hole horizon,
which means, $0< z\ll 1$. Now rotate $z$ in the complex $z$ plane
over an angle $2\pi $ clockwise and examine what happens to the wave
function. The validity of semiclassical
wave function is justified by keeping $\mid z\mid \ll 1$%
. This analytical continuation necessarily incorporates a crossing
of the cut on the complex plane. Therefore, after finishing this
rotation and returning to a real physical value $z>0$, the wave
function acquires a new value on
its Riemannian surface. Let it be $\Phi ^{2\pi }_{l}\left( r\right) $. Then,%
\begin{equation}
\label{21}\Phi ^{2\pi }_{l}\left( r\right) =\rho \exp \left( -\imath
\xi \ln \left( r-r_{+}\right) \right) +\frac{\mid R\mid }\rho \exp
\left( +\imath \xi \ln \left( r-r_{+}\right) \right),
\end{equation}
 where
$\rho =\exp \left( -2\pi \xi \right) $. The analytically continued
function $\Phi ^{2\pi }_{l}\left( r\right) $ satisfies the same
differential equation as the initial wave function $\Phi _{l}\left(
r\right)$. And one has to expect that $\Phi ^{2\pi }_{l}\left(
r\right) $ must satisfy the same normalization condition as the
initial wave function $\Phi _{l}\left(r\right) $.
 This implies that one of the coefficients, either $\rho $ or $%
\frac{\mid R\mid }\rho $ should have an absolute value equal to unity. Since $%
\rho <1$, we assume that $\frac{\mid R\mid }\rho =1$, thus $R=\exp
\left( -2\pi \xi \right) $. We see that reflection coefficient is
non zero. In other words, black hole horizon is capable of
reflection. The probability of reflection from the horizon can be
found as $P=\mid R\mid ^2=$ $\exp \left( -4\pi \xi \right) $. Since
the reflection is taking place against the background of a  black
hole with temperature $T$, we see that $P=\exp \left( -\frac{\left( \epsilon -eA_t\right) }%
T\right) $, where $\epsilon$ is the energy of the particles, $e$ the
electric charge and $A_t=\frac qr$  is the electric potential.
Therefore the Hawking temperature of RN black hole is ,
\begin{equation}
\label{z6}T=\frac{r_{+}-r_{-}}{4\pi r_{+}^2}.
\end{equation}

\subsection{Region 2: Intermediate region: $r>r_{+}$ }

The region is sufficiently away from $r_{+}$ but not very far away
from $r_{+}$ . Here the terms in $( \epsilon -eA_t) ^2$ and $\mu ^2$
are much smaller than all other terms. Thus we neglect the low
energy and momentum in Eq. (\ref{h6}), and for s-wave it becomes,
\begin{equation}
\Phi _0^{\prime \prime }(r)+\frac{2r-1}{r^2-r+q^2}\Phi _0^{\prime
}(r)=0.
\end{equation}
Therefore $ \ln \Phi _0^{\prime }(r)=-\ln \left( r^2-r+q^2\right)
+\ln C $ and it can be written as,
\begin{equation}
\Phi _0^{\prime }(r)=\frac C{r^2-r+q^2}=\frac A{r-r_{+}}+\frac
B{r-r_{-}}.
\end{equation}
Since $r>r_{+\text{ }}$ we neglect the effect of $r_{-}$, and the
above equation can be written as,
\begin{equation}
\Phi _0(r)=\int \frac K{\left( r-r_{+}\right) r}dr,
\end{equation}
i.e.,
\begin{equation}
\label{m}\Phi _0(r)=\alpha \ln \frac{\left( r-r_{+}\right) }r+\beta.
\end{equation}
\subsubsection{Comparing regions 1 and 2}
The definitions of the above two regions do not really lead to any
overlap region. However near the point $r_{+}$ one can approximate
the solutions by linear combinations of constant terms and terms
proportional to $\ln(r-r_{+}) $. To obtain this consider Eq. (\ref{p1}) which is the wave function in the region $%
r\rightarrow r_{+}$%
\begin{equation}
\Phi_{l} \left( r\right) =\exp \left( -\imath \xi \ln \left(
r-r_{+}\right) \right) +\mid R\mid \exp \left( +\imath \xi \ln
\left( r-r_{+}\right) \right).
\end{equation}
We can take $\exp \left( \pm \imath \xi \ln \left( r-r_{+}\right)
\right) =1\pm \left( \imath \xi \ln \left( r-r_{+}\right) \right) $,
therefore the above equation becomes,
\begin{equation}
\label{q6}\Phi \left( r\right) =1+\mid R\mid -\left( 1-\mid R\mid
\right) \imath \xi \ln \left( r-r_{+}\right).
\end{equation}
Eq. (\ref{m}) can be written as,
\begin{equation}
\label{m6}\Phi _0(r)=\alpha \ln \left( r-r_{+}\right) +\beta.
\end{equation}
Comparing Eq. (\ref{m6}) with Eq. (\ref{q6}) we get,
\begin{equation}
\label{r}\alpha =-\imath \xi \left( 1-\mid R\mid \right) \text{,
}\beta =1+\mid R\mid.
\end{equation}
\subsection{Region 3: Far away from the horizon: $ r>>r_{+}$}
Now in the region $ r>>r_{+}$ we can write $2r-1=2r-\left(
r_{+}+r_{-}\right) =\left( r-r_{+}\right) +\left( r-r_{-}\right)$.
Therefore Eq. (\ref{h6}) becomes:
\begin{eqnarray}
\label{k6}\Phi _l^{\prime \prime }(r)+\left( \frac 1{r-r_{+}}+\frac
1{r-r_{-}}\right) \Phi _l^{\prime }(r)\\ \nonumber+\left(
\frac{\left( \epsilon
-eA_t\right) ^2r^4}{\left( r^2-r+q^2\right) ^2}-\frac{\mu ^2r^2}{r^2-r+q^2}-%
\frac{l\left( l+1\right) }{r^2-r+q^2}\right) \Phi _l(r)=0.
\end{eqnarray}
In the above equation the terms containing energy and mass  can be
simplified as,
\begin{eqnarray}
\label{f}\frac{\left( \epsilon -eA_t \right) ^2r^4}{\left(
r^2-r+q^2\right)
^2}=\left( \epsilon -eA_t\right) ^2+\frac{2\left( \epsilon -eA_t\right) ^2r%
}{\left( r^2-r+q^2\right) }+\\ \nonumber\frac{\left( \epsilon
-\frac{eq}r\right) ^2\left( r^2-2r^2q^2-q^4\right) ^{}}{\left(
r^2-r+q^2\right) ^2},
\end{eqnarray}
and
\begin{equation}
\label{g}\frac{\mu ^2r^2}{r^2-r+q^2}=\mu ^2+\frac{\mu ^2r}{r^2-r+q^2}-\frac{%
\mu ^2q^2}{r^2-r+q^2}.
\end{equation}
Thus,
\begin{eqnarray}
\frac{\left( \epsilon -eA_t\right) ^2r^4}{\left( r^2-r+q^2\right) ^2}-\frac{%
\mu ^2r^2}{r^2-r+q^2}=\left( \epsilon -eA_t\right) ^2-\mu ^2+\\
\nonumber\frac{\left( 2\left( \epsilon -eA_t\right) ^2-\mu ^2\right)
r}{\left( r^2-r+q^2\right) },
\end{eqnarray}
\begin{equation}
\label{h1}=p^2+\frac{\left( p^2+\left( \epsilon -eA_t\right) ^2\right) r}{%
\left( r^2-r+q^2\right) },
\end{equation}
where $p$ is the momentum and is given by $p^2=\left( \epsilon
-eA_t\right) ^2-\mu ^2$. Since $r$ is very large, only terms up to
$\frac 1r$ is taken. Substituting Eq. (\ref{h1}) in Eq. (\ref{k6})
we get,
\begin{eqnarray}
\label{j}\Phi _l^{\prime \prime }(r)+\frac 2r\Phi _l^{\prime }(r)+\\
\nonumber\left( p^2+ \frac{\left( p^2+\left( \epsilon -eA_t\right)
^2\right) }{2r}-\frac{l\left( l+1\right) }{r^2}\right) \Phi _l(r)=0.
\end{eqnarray}
This equation is of Coulomb type where the Coulomb charge is
$Z=\frac{\left( \epsilon -eA_t\right) ^2+p^2}2$ and the solution to
this equation can be written,
\begin{equation}
\label{k}\Phi _l(r)=\frac 1r\left( A_l\exp \left( \imath z\right)
+B_l\exp \left( -\imath z\right) \right),
\end{equation}
where $z=pr-\frac{l\pi }2+\nu \ln 2pr+\delta _t^{(c)}$, where
$\delta _t^{(c)}=\arg \Gamma \left( l+1-\nu \right)$, where $\nu
=\frac Zp$ \cite{ee}. In the region of large separations $r\gg r_+$.
 The Coulomb wave function has a regular singularity at
$r=0$ and it has an irregular singularity at $r=\infty$. Let $%
F_{l}\left( r\right) $ be the regular Coulomb wave function and
$G_{l}\left( r\right) $ be the irregular Coulomb wave function. Then
the solution of Coulomb problem can be presented as a linear
combination of these two functions:
\begin{equation}
\label{s}\Phi _{l}(r)=\frac 1r\left( aF_{l}\left( r\right)
+bG_{l}\left( r\right) \right).
\end{equation}
In the asymptotic region $r\rightarrow \infty $, for Coulomb
functions we can use the known formulae, $ F_{l}\left( r\right)
=\sin z\text{, }G_{l}\left( r\right) =\cos z $ where
$z=pr-\frac{l\pi }2+\nu \ln 2pr+\delta _t^{(c)}$, thus Eq. (\ref{s})
will be in an asymptotic form:
\begin{equation}
\label{11}\Phi _{l}(r)=\frac 1r\left( a\sin z\ +b\cos z\ \right).
\end{equation}
But we know that, for $l=0$, \cite{ee}
\begin{equation}
F_{0}\left( r\right) =cpr\,\text{, }G_{0}\left( r\right) =\frac 1c,
\end{equation}
where
\begin{equation}
c^2=\frac{2\pi \nu }{1-\exp \left( 2\pi \nu \right) }.
\end{equation}
Thus Eq. (\ref{s}) for s wave will be,
\begin{equation}
\label{s1}\Phi_{0} (r)=acp+\frac b{cr}.
\end{equation}
\subsubsection{Comparing Regions 2 and 3}
To compare the wave function in the regions 2 and 3, we write Eq.
(\ref{m}) as,
\begin{equation}
\label{t}\Phi _0(r)=\alpha \ln \left( 1-\frac{r_{+}}r\right) +\beta \simeq -%
\frac{\alpha r_{+}}r+\beta,
\end{equation}
since $\ln \left( 1-\frac{r_{+}}r\right) =-\frac{r_{+}}r$. Thus from
Eq. (\ref{s1}) and Eq.~ (\ref{t}) we get,
\begin{equation}
\label{u}\text{ }a=\frac{1+\mid R\mid }{pc}\text{, }b=\imath \xi
r_{+}c\left( 1-\mid R\mid \right) .
\end{equation}
\section{absorption cross section}
Now we will find an expression for the absorption cross section of
RN black hole. The two terms in Eq. (\ref{k}) represent the incoming
and outgoing waves. The $S$ matrix can be written as the ratio of
coefficient of the incoming and outgoing waves. Therefore,
\begin{equation}
\label{l}S_l=\left( -1\right) ^{l+1}\frac{A_l}{B_l}\exp \left(
2\imath \delta_{l}\right),
\end{equation}
where $A_l$ represents the amplitude of the incident wave and $B_l$
that of the reflected wave. We have to find $S$ matrix. Since the
latter decreases exponentially with energy we will consider first
the low energy region $\epsilon \ll 1$, where reflection from the
horizon is prominent and restricted to $l=0$ and denoting s wave as
$\Phi _l(r)=\Phi _0\left( r\right) $.
Thus we can find coefficient $%
A_0$, $B_0$ in the latter. Using Eq. (\ref{11}) we can deduce,
\begin{equation}
\label{v}A_0=\frac{a+\imath b}{2\imath }\text{, }B_0=\frac{-a+\imath b}{%
2\imath }.
\end{equation}
Employing Eq. (\ref{u}) we can find,
\begin{equation}
A_0=\frac{\left[ 1+\mid R\mid -\xi c^2p\left( 1-\mid R\mid \right)
r_{+}\right] }{2\imath pc},
\end{equation}
and
\begin{equation}
B_0=-\frac{\left[ 1+\mid R\mid +\xi c^2p\left( 1-\mid R\mid \right)
r_{+}\right] }{2\imath pc}.
\end{equation}
Corresponding S-matrix from Eq. (\ref{l}) for the s-wave is given
by,
\begin{equation}
S_0=-\frac{A_0}{B_0}\exp \left( 2\imath\delta_{0} \right)
=\frac{1+\mid R\mid -\xi c^2p\left( 1-\mid R\mid \right)
r_{+}}{1+\mid R\mid +\xi c^2p\left( 1-\mid R\mid \right) r_{+}}\exp
\left( 2\imath \delta_{0}\right),
\end{equation}
which can be written as,
\begin{equation}
\label{w}S_0=\frac{1-\xi c^2pr_{+}\eta }{1+\xi c^2pr_{+}\eta }\exp
\left( 2\imath \delta_{0}\right),
\end{equation}
where $ \eta =\frac{1-\mid R\mid }{1+\mid R\mid } $. The absorption
cross section in the low energy limit is given by,
\begin{equation}
\label{x}\sigma _{abs}=\frac \pi {p^2}\left( 1-\mid S_o\mid
^2\right) =\frac \pi {p^2}\frac{4c^2\xi pr_{+}\eta }{\left( 1+\xi
c^2pr_{+}\eta \right) ^2}.
\end{equation}
Taking $p=\left( \epsilon -eA_t\right) v$, we write Eq. (\ref{x})
as,
\begin{equation}
\label{y}\sigma _{abs}=\frac{4\pi c^2r_{+}^3\eta }{v\left(
r_{+}-r_{-}\right) \left( 1+\xi c^2pr_{+}\eta \right) ^2}.
\end{equation}
We know that Hawking temperature is given by,
\begin{equation}
\label{z6}T=\frac{r_{+}-r_{-}}{4\pi r_{+}^2}.
\end{equation}
Therefore
\begin{equation}
\label{y6}\sigma _{abs}=\frac{c^2r_{+}\eta }{vT\left( 1+\xi
c^2pr_{+}\eta \right) ^2}.
\end{equation}

\begin{figure}
\includegraphics[width=7cm]{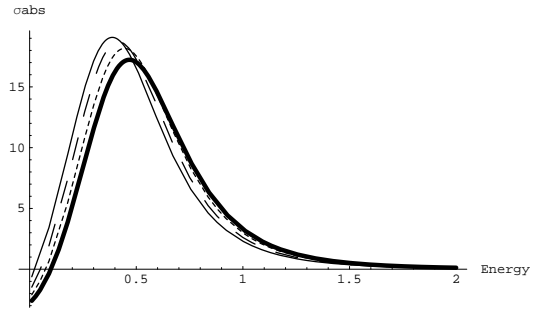}
\caption{$\sigma_{abs}$ versus $\epsilon$ for RN black hole with
reflection, is plotted for different charges. The solid curve is for
$q=0.1$, dashed curve is for $q=0.2$, dotted curve is for $q=0.3$,
and bold curve is for $q=0.4$.}\label{graph1}
\end{figure}

In the expression for absorption cross section we have to substitute
for $c^{2}$ and $\eta$. We know that $\nu =\frac Zp$, where
$Z=\frac{\left( \epsilon -eA_t\right) ^2+p^2}2$ and therefore $ \nu
=\frac{\left( \epsilon -eA_t\right) ^2+p^2}{2p}=\frac{\left(
\epsilon -eA_t\right) v}2\left( 1+\frac 1{v^2}\right)$. Hence,
\begin{equation}
c^2=\frac{\pi \left( \epsilon -eA_t\right) v\left( 1+\frac 1{v^2}\right) }{%
1-\exp \left( -\pi \left( \epsilon -eA_t\right) v\left( 1+\frac
1{v^2}\right) \right) },
\end{equation}
for low energy,
\begin{equation}
\label{qm23}c^2=\frac{\pi \left( \epsilon -eA_t\right) v\left(
1+\frac 1{v^2}\right) }{1-(1-\pi \left( \epsilon -eA_t\right)
v\left( 1+\frac 1{v^2}\right) )}\simeq 1,
\end{equation}
and
\begin{equation}
\label{qm24}\eta =\frac{1-\exp \left( -2\pi \xi \right) }{1+\exp
\left( -2\pi \xi \right) }=\tanh \pi \xi \simeq \pi \xi.
\end{equation}
Substituting Eqs. (\ref{qm23}) and (\ref{qm24}) in Eq.~(\ref{y6}) we
find,
\begin{equation}
\label{ans24}\sigma _{abs}=\frac{r_{+}\pi \xi }{vT\left(
1+\frac{\left( \epsilon -eA_t\right) ^2r_{+}}{16\pi ^2T^2}\right)
^2}\simeq \frac{\left( \epsilon -eA_t\right) r_{+}}{4vT^2}.
\end{equation}
i.e, the absorption cross section of RN black hole is found to
depend inversely on square of Hawking temperature. Now if $q=0$, the
metric becomes of Schwarzschild type. Thus if we substitute $T=\frac
1{4\pi r}$ and $r_{+}=r$ we will get Kuchiev's result \cite{ff}
\begin{equation}
\label{ans25}\sigma _{abs}=\frac{4\pi ^2r^3\epsilon }v,
\end{equation}
and in the absence of reflection we arrive at the result obtained by
Unruh \cite{pp}. Now using Eq. (\ref{y}) we plot $\sigma _{abs}$
versus $\varepsilon$. The curve is plotted for RN black hole with
reflection and taking charges q=0.1, 0.2, 0.3, 0.4. And we found
that absorption cross section decreases when charge is increased
from 0.1 to 0.4, i.e., there is more possibility of reflection. It
is shown in Figure \ref{graph1}.

\begin{figure}
\includegraphics[width=7cm]{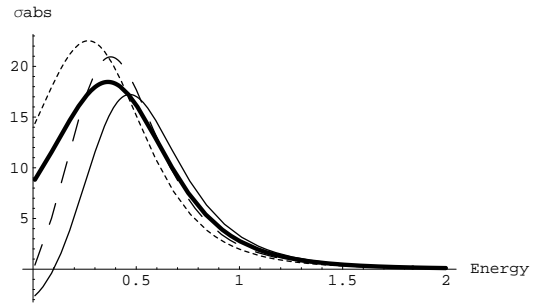}
\caption{ $\sigma _{abs}$ versus $\varepsilon$ for RN black hole
with (solid curve) and without (bold curve) reflection and for
Schwarzschild black hole with (dashed curve) and without reflection
(dotted curve).}\label{graph2}
\end{figure}

In Figure \ref{graph2} we plot $\sigma _{abs}$ versus $\varepsilon$
for RN black hole with and without reflection and for Schwarzschild
black hole with and without reflection. From the plot it is clear
that absorption cross section is decreased by the presence of
charge. And also found that for RN case the graph is shifted to
right.

\section{conclusion}
We  found the wave function $\Phi_{l} (r)$ in the vicinity of outer
horizon of RN black hole i.e $r\rightarrow r_+$ for charged scalar
field using WKB approximation. We have also studied the behavior of
scattered charged scalar waves in the regions $r>r_+$ and $r\gg r_+$
in low energy limit. By comparing the solutions in the 3 regions
viz., $r\rightarrow r_+$, $r>r_+$ and $r\gg r_+$, we found the
S-matrix and the absorption cross section for RN black hole in the
lower energy limit. The absorption cross section is found to be
inversely depending on the square of the Hawking temperature. From
$\sigma _{abs}$ of RN, we deduced the absorption cross section of
Schwarzschild black hole in the presence of reflection and in the
absence of reflection, which agree with the results obtained earlier
\cite{ff,pp}. By plotting $\sigma _{abs}$ versus $\varepsilon$ plot
it is found that absorption cross section is decreased by increasing
the of charge in RN black hole.

\section{Acknowledgement}

 SR is thankful to Cochin University of Science and Technology for financial support in the form of University
Senior Research Fellowship. VCK wishes to acknowledge Associateship
of IUCAA, Pune and U.G.C for financial support in the form of a
Major Research Project.


\begin{thebibliography}{0}
\bibitem{ss} J.A.H. Futterman, F.A. Handler and R.A. Matzner, {\it Scattering from Black holes} (Cambridge University Press, Cambridge,
1988).
\bibitem{tt} J. Hough and S. Rowen,  Living Rev.Rel. \textbf{3}, 3
(2000).
\bibitem{ii}  R. Penrose, Riv. Nuovo Cimento  \textbf{1}, 252
(1969).
\bibitem{abaa} S.W. Hawking, Commun. Math. Phys.\textbf{43}, 199(1975).
\bibitem{pp}  W. G. Unruh,  \prd  \textbf{14}, 3251 (1976).
\bibitem{abab} N. Sanchez, \prd \textbf{16}, 937 (1977).
\bibitem{abac} N. Sanchez, \prd \textbf{18}, 1030 (1978).
\bibitem{abad} N. Sanchez, \prd \textbf{18}, 1798 (1978).
\bibitem{viv} G. 't Hooft,  gr-qc/0406017.
\bibitem{22} Mu-Lin Yan and Hua Bai, gr-qc/0401027.
\bibitem{bb}  M.Yu. Kuchiev, \prd  \textbf{69}, 124031 (2004).
\bibitem{sr}  R. Sini and V.C. Kuriakose, Int. J. Mod. Phys D \textbf{16}, 105 (2007).

\bibitem{uu} B.F. Schutz and C.M. Will, Astrophys. J. Lett.  \textbf{291}, L33
(1985).
\bibitem{vv} S. Iyer and C.M. Will,  \prd\textbf{35}, 3621 (1987).
\bibitem{bn} S.W. Hawking and G.F.R. Ellis, {\it The large scale structure of space-time} (Cambridge, University Press,
Cambridge, 1973).

\bibitem{ee}  L. D. Landau and E. M. Lifshits, {\it Quantum Mechanics:
Non-Relativistic Theory} (Pergamon, New York, 1977).

\bibitem{ff}  M.Yu. Kuchiev, V.V. Flambaum, \prd \textbf{70}, 044022
(2004).


\end{thebibliography}
\end{document}